\documentclass[prx,superscriptaddress,floatfix,nofootinbib,a4paper,twocolumn,aps,10pt]{revtex4-1}
\usepackage[utf8]{inputenc}
\usepackage{natbib}
\usepackage{amsmath}
\usepackage{amsbsy}
\usepackage{amssymb}
\usepackage{scrextend}
\usepackage{graphicx}
\usepackage{color}
\usepackage{array}
\usepackage[usenames, dvipsnames]{xcolor}
\usepackage[colorlinks=true, allcolors = WildStrawberry]{hyperref}
\graphicspath{{graphics/}}
\renewcommand{\vec}{\boldsymbol}
\newcommand{\kB}{k_{\scriptscriptstyle B}}

\newcommand{\mNoBracket}[1]{\begin{matrix} #1 \end{matrix}}

\DeclareMathOperator{\Erf}{Erf}

\begin{document}
\title{Statistical stability of random potentials to thermal and quantum activation}
\author{L.\ Filsinger}
\affiliation{Institute for Theory of Condensed Matter, Karlsruhe Institute of
Technology, 76131 Karlsruhe, Germany}
\author{R.\ Willa}
\affiliation{Institute for Theory of Condensed Matter, Karlsruhe Institute of
Technology, 76131 Karlsruhe, Germany}
\affiliation{Heidelberger Akademie der Wissenschaften, 69117 Heidelberg, Germany}
\affiliation{Institute of System Engineering, University of Applied Sciences, 1950 Sion, Switzerland}
\date{\today}

\begin{abstract}
In numerous physical, chemical, and biological systems the dynamics can be reduced to the motion of state variables in a complex potential landscape. In case the manifold is known, the motion and response of the embedded object can be described deterministically up to stochastic effects usually associated with a noise. In contrast, if the manifold is unknown, the static and dynamic response of the state variable may be used as a spectroscopic tool to characterize the potential landscape. Inspired by a seminal work of L.\ Embon and co-workers, [Sci.\ Rep.\ \textbf{5}, 7598 (2015)] we investigate the statistical properties of potential minima, in particular, their stability to thermal and quantum activation. For Gaussian random manifolds, we derive an algebraic expression to evaluate the statistical probability of the potential character (value, slope, curvature, ...). With this tool, we compute the expectation value for the rate of thermal and quantum activation and link these findings to the principal characteristics of the Gaussian potential, i.e., its Green's function. This link provides the opportunity to access information on the potential's Green's function by studying the activation behavior of an object in this manifold.
\end{abstract}

\maketitle

\section{Introduction}

Numerous physical systems are subject to randomness or disorder \cite{Nattermann1990, Guinier1994, Bertet2005, Ragan2009, Stern2013, Gu2021} and studying the properties of random noise has triggered a number of mathematical works \cite{Kac1943, Rice1945, Kac1948, LonguetHiggins1957, LonguetHiggins1960,Fyodorov2015}. The precise microscopic response of the system usually remains inaccessible. At the same time, the global (macroscopic) response of a large collection of objects subject to disorder often follows a deterministic behavior, owing to the fact that the randomness averages out over many realizations and yields an effective response. The same applies if an individual experiment is repeated multiple times and the response function is averaged over many realizations.
We are motivated by the particular situation of superconducting vortices (quantized flux lines) moving in a disorder landscape provided by material defects. Characterizing the macroscopic dynamics of vortices and the optimization for technological applications has been the subject of decades-long research endeavors. A leap in scanning imaging techniques (STM, SQUID-on-Tip) now allows to microscopically study the motion of these vortices in their pinning landscape (or simply pinscape) and to study its character \cite{Guillamon2011, Embon2015, Kremen2016, Herrera2017, Embon2017, Willa2020a}.
Theoretically, the precise knowledge of the system's randomness allows to compute effective responses by statistical average. Reversing the argument, one may use the system's effective response to characterize the underlying randomness. In this manuscript, we want to tackle this inverse problem. 

To make the statement more specific, let us consider a single particle (e.g.\ vortex) in a metastable minimum of a random pinscape. The escape of the particle from this minimum---for instance through external forces, thermal fluctuations, or quantum tunneling---clearly depends on the specific properties of the pinning potential. On the other hand, the average escape rate over many single-particle-in-a-trap realizations becomes predictable and can serve as a spectroscopic tool to learn about the potential landscape.

This work provides two steps towards pinscape spectroscopy: As a first step (Sec.~\ref{sec:statistical-to-algebraic}) we demonstrate that the statistical average of observables can be reduced to an algebraic one and illustrate this step for the widely-used class of Gaussian random potentials. This generic result is obtained by computing the full probability distribution for the Taylor coefficients of the potential landscape. In a second step (Sec.~\ref{sec:results}) we compute several macroscopic observables; most notably the escape rate from a potential minimum by thermal and quantum activation and their response to temperature and particle mass, respectively. These observables are directly linked to the characteristics of the potential and hence serve as spectroscopic lenses to quantify the system's disorder. The Gaussian statistics used in these analytic approaches are reached as a central limit when disorder is produced by an infinite density of defects. Aiming for a realistic description, we provide in Sec.~\ref{sec:numerics} a numerical tool to compute the correlation function at finite densities.

\section{Reduction of statistical problem to algebraic one}\label{sec:statistical-to-algebraic}
We consider a point-like particle in a potential landscape $U(x)$, with $x$ denoting the spatial coordinate. While the results can be derived in arbitrary dimensions, we will consider the one-dimensional case here. For a smooth manifold the potential can be expanded in the vicinity of any location $x_{0}$ as
\begin{align}\label{}
   U(x) &\approx u_{0} + \sum\nolimits_{n=1}^{\infty} \frac{u_{n}}{n!} (x-x_{0})^{n}.
\end{align}
The expansion coefficients $u_{n} = U^{(n)}(x_{0})$ denote the $n$-th derivative of the potential at $x_{0}$. For the particle to be at rest in a stable minimum at $x_{0}$, it must hold that $u_{1} = 0$ and $u_{2}>0$.
Often the third derivative is dominating the proximity of a stable point to the next instability, see discussion in Appendix \ref{app:truncation}. In fact truncating the expansion at $n = 3$ the barrier to escape from the potential trap is
\begin{align}\label{eq:threshold}
   U_{b} = 2 u_{2}^{3} / 3 u_{3}^{2}
\end{align}
This tells, that if the system is subject to a destabilizing perturbation of strength $E$, the particle at such a position remains stable if $E < U_{b}$ and moves to a neighboring minimum otherwise. This consideration raises the question what is the statistical relation between $u_{2}$ and $u_{3}$ for realizing a confining potential with $U_{b} < E$. This line of thought can be extended to the activation over the barrier through thermal fluctuations---the activation frequency follows Arrhenius law $\Omega = \omega_{0} e^{-U_{b}/\kB T}$---or the quantum tunneling of the particle through the barrier with a tunneling probability $|\mathcal{T}|^{2} \sim e^{-2\Lambda_{b}}$
, see below.

Having this question in mind, we deploy a statistical method to determine the probability distribution $p(\{u_{n}\})$ that a subset $\{U^{(n)}(x_{0})\}$ of the expansion parameters assume the specific values $\{u_{n}\}$. When appropriate the sets $\{\cdot\}$ are brought in a compact form with the vector notations $\vec{n} \equiv \{n\}$, $\vec{u} \equiv \{u_{n}\}$, \dots , and $N$ the dimension of these vectors. Rather than averaging over the position $x_{0}$, it is more convenient---and mathematically equivalent as shown in Ref.~\cite{Willa2022}---to opt for a statistical average over \emph{potential realizations}. The probability $p(\vec{u})$ can then formally be evaluated from
\begin{align}\label{eq:probability-definition}
   p(\vec{u}) &= \!\!\int \!\! \mathcal{D}[U(r)] \, 
   \mathcal{P}[U(r)] 
   \prod\nolimits_{\vec{n}}\delta[U^{(n)}(0) \!-\! u_{n}],
\end{align}
where $\int \!\! \mathcal{D}[U(r)] \mathcal{P}[U(r)]$ denotes the functional average with $\mathcal{P}[U(r)]$ the functional probability distribution for realizing $U(r)$ and $\delta(u)$ is the Dirac delta distribution. Also, $x_{0} = 0$ has been chosen without loss of generality.
Knowing $p(\vec{u})$, the statistical realization average
\begin{align}\label{eq:statistical-average}
   \langle f \rangle \equiv \!\!\int \!\! \mathcal{D}[U(r)] \, \mathcal{P}[U(r)] f
\end{align}
of any function $f(\vec{u})$ reduces to an algebraic problem
\begin{align}\label{eq:algebraic-average}
   \langle f \rangle = \int \prod\nolimits_{\vec{n}} du_{n} p(\vec{u}) f(\vec{u}).
\end{align}
Let us highlight here, that Eq.~\eqref{eq:statistical-average} involves a \emph{functional integration}. Instead, Eq.~\eqref{eq:algebraic-average} is an \emph{algebraic integral} over as many degrees of freedom, as the function $f$ depends on.
We apply this complexity reduction scheme on the class of Gaussian random potentials with vanishing
mean $\langle U(x) \rangle = 0$ and a two-point correlation function
\begin{align}\label{eq:correlator}
   \!\!
   G(x - y) \!\equiv\! \langle U(x) U(y) \rangle,
\end{align}
This type of potential landscape is characterized by a Gaussian functional distribution
\begin{align}\label{eq:Gaussian-measure}
   \mathcal{P}[U(r)] = 
   e^{-\mathcal{S}} / \mathcal{Z},
\end{align}
where the action
\begin{align}\label{eq:Gaussian-action}
   \mathcal{S} = \frac{1}{2} \int\! \frac{dx}{L}\! \int\!
   \frac{dy}{L}\ U(x) G^{-1}(x-y) U(y)
\end{align}
is a quadratic form in $U$ and $\mathcal{Z} = \int \mathcal{D}[U(r)]\ e^{-\mathcal{S}}$ is the partition sum. Here, $L$ denotes the system size. This type of random potential is commonly used in statistical and computational sciences \cite{SaYakanit1979,Peacock1999,Hanson2010,Dobramysl2014}. As derived in Ref.~\cite{Willa2022}, a specific realization of such a Gaussian distribution is reached for a large density $N_{p}/L \gg 1$ of smooth potential wells $V(x - x_{\ell})$ randomly distributed at positions $x_{\ell}$, i.e. for
\begin{align}\label{}
   U(x) = \sum\nolimits_{\ell = 1}^{N_{p}} V(x - x_{\ell}).
\end{align}
In this example, the two-point correlator takes the explicit form $G(x-y) =  (N_{p}/L) \int dr V(x-r) V(y-r)$ and can be evaluated from the shape of the constituting wells.

We now proceed with evaluating the probability in Eq.~\eqref{eq:probability-definition} for the Gaussian measure, see Eqs.~\eqref{eq:Gaussian-measure} and \eqref{eq:Gaussian-action}. Expressing each $\delta$-distribution in Fourier space we arrive at
\begin{align}\label{}
   \!\!\!
   p(\vec{u}) &= \!\int\! \mathcal{D}[U(r)] \frac{e^{-\mathcal{S}}}{\mathcal{Z}}
                    \prod_{\vec{n}} \!\Big[\!\int\! \frac{dk_{n}}{2\pi}e^{i k_{n} [u_{n} - U^{(n)}(0)] }\Big]\!\!
\end{align}
Rewriting
\begin{align}\label{}
   U^{(n)}(0) = \frac{1}{2} \int \frac{dx}{L}\frac{dy}{L} [U^{(n)}(x) \delta(x) + U^{(n)}(y) \delta(y)]
\end{align}
and applying $n$ integrations by parts to the above expression we arrive at
\begin{align}\label{}
   p(\vec{u}) &=\! \int \frac{d\vec{k}}{(2\pi)^{N}} e^{i \vec{k} \cdot \vec{u}}
   \frac{1}{\mathcal{Z}} \int \mathcal{D}[U(r)] e^{-\mathcal{S}} \times
   \\\nonumber
   &\qquad
   e^{-\frac{1}{2} \int \frac{dx}{L}\frac{dy}{L} \sum_{\vec{m}} (-1)^{m} i k_{m} [U(x) \delta^{(m)}(x) + U(y) \delta^{(m)}(y)] },
\end{align}
with $\delta^{(m)}$ denoting the $m$-th derivative of the delta distribution. As $\mathcal{S}$ possesses a quadratic form, this functional integral can be solved
, yielding
\begin{align}\label{}
   p(\vec{u}) &=\! \int \frac{d\vec{k}}{(2\pi)^{N}} e^{i \vec{k} \cdot \vec{u}} \times
   \\\nonumber
   &\qquad\quad
   e^{-\frac{1}{2} \!\int\! \frac{dx}{L}\frac{dy}{L} \sum_{\vec{m}, \vec{\ell}} (-1)^{m + \ell} k_{m} k_{\ell} \delta^{(m)}(x)\delta^{(\ell)}(y) G(x-y) }.
\end{align}
Another series of integration by parts provides us with
\begin{align}\label{}
   p(\vec{u}) &= \int \frac{d\vec{k}}{(2\pi)^{N}} e^{i \vec{k} \cdot \vec{u}}
   e^{-\frac{1}{2} \sum_{\vec{m}, \vec{\ell}} (-1)^{\ell} k_{m} k_{\ell} G^{(m+\ell)}},
\end{align}
with $G^{(m+\ell)} \equiv G^{(m+\ell)}(0)$ the $(m \!+\! \ell)$-th derivative of the two-point correlator at the origin, i.e., for $x = 0$. It generically holds that odd derivatives vanish as $G(r) \!=\! G(-r)$ is an even function in $r$. Furthermore it holds that $(-1)^{\ell}G^{(2\ell)} > 0$, that is, the sign of (even) derivatives alternates. 
Defining the matrix $\mathcal{G}_{m\ell} \equiv (-1)^{\ell} G^{(m+\ell)}$, we arrive at
\begin{align}\label{eq:general-result}
   p(\vec{u}) &= \frac{(2\pi)^{-N/2}}{\sqrt{\det(\mathcal{G})}}
   e^{-\frac{1}{2} \vec{u} \mathcal{G}^{-1} \vec{u} }.
\end{align}
This expression constitutes the key finding of the section. Before proceeding with applying this result to a physical problem, let us highlight few properties of this probability density and discuss some special cases. As $G^{(2\ell+1)} = 0$ for all integers $\ell \geq 0$, it immediately follows that the probabilities of even and odd derivatives do not interfere with one another. Specifically one can decompose $p(\vec{u}) = p(\vec{u}_{e})p(\vec{u}_{o})$ into a product of probability densities of the even/odd derivatives $p(\vec{u}_{e/o})$, respectively. In other words, values of even derivatives in $U(x)$ are uncorrelated to those of odd derivatives and vice-versa. For the potential's $n$-th derivative, the above general formula \eqref{eq:general-result} assumes the Gaussian form
\begin{align}\label{}
   p(u_{n}) = e^{- u_{n}^{2} / [2|G^{(2n)}|]} / \big(2\pi |G^{(2n)}|\big)^{1/2},
\end{align}
where the $2n$-th derivative of the two-point correlator is a measure of the characteristic distribution width. For the $n$-th and $m$-th derivatives where $n+m$ is odd it holds that $p(u_{n},u_{m}) = p(u_{n}) p(u_{m})$ as discussed earlier. If $n+m$ is even we have 
\begin{align}\label{}
   \!\!
   p(u_{n},u_{m}) &=\\\nonumber
   &\frac{\exp\!\Big(\!\! -\!\frac{|G^{(2m)}|u_{n}^{2} -  2 (-1)^{n} G^{(n+m)}u_{n} u_{m} + |G^{(2n)}|u_{m}^{2}}
      {2 [ G^{(2n)} G^{(2m)} - (G^{(n+m)})^{2}]}\Big)}{2\pi \sqrt{G^{(2n)} G^{(2m)} - (G^{(n+m)})^{2}}}.
\end{align}
In the following, we compute observables associated with various derivatives of the potential landscape by using these probability densities.

\section{Results}\label{sec:results}
By virtue of Eqs.~\eqref{eq:algebraic-average} and \eqref{eq:general-result}, the statistical computation \eqref{eq:statistical-average} of observable quantities has now become an algebraic problem only involving those derivatives $\vec{u} = \{u_{n}\}$ which affect the function $f$. In the following we apply this concept to different observable quantities: Special area fractions, thermal activation rates and quantum tunneling rates.

\subsection{Area fractions}\label{sec:area-fractions}
A particularly simple example to start with is the computation of \emph{area fractions}, i.e. the fraction of points satisfying certain criteria. For instance the probability to find a position where the potential assumes a negative value $u_{0} < 0$ is formally given by
\begin{align}\label{}
   \mathfrak{p}_{[u_{0}<0]} &= \!\int_{-\infty}^{0} \!\!\!du_{0} \Big[\prod_{n=1}^{\infty} \int_{-\infty}^{\infty} \!\!\!du_{n}\Big] p(\{u_{0}, u_{1}, u_{2}, \dots\})\\
   &= \int_{-\infty}^{0} \!\!\!du_{0} p(u_{0}) = 1/2.
\end{align}
The fraktur typesetting $\mathfrak{p}$ distinguishes probabilities from the probability densities $p$. Furthermore, the constraint to which the calculated probability is subjected to is specified as a subscript. For any specific derivative $u_{j}$, the area fraction of finding it with a definite sign is 1/2, i.e., $\mathfrak{p}_{[u_{j} > 0]} = \mathfrak{p}_{[u_{j} < 0]} = 1/2$. This property of the Gaussian random potential holds for a density of pinning wells only in the central limit $N_{p}/L \to \infty$.
For pinning problems, the \emph{stable area fraction}, defined in Ref.~\cite{Willa2022} as the fraction of points that become a minimum upon appropriate tilt (mathematically $\mathfrak{p}_{[u_{2} > 0]}$) amounts to 1/2 in one dimension. Let us highlight here, that in two and higher dimensions the stable area fraction (points with a positive definite Hessian matrix) is $(3-\sqrt{3})/6 \approx 0.21 < 1/2$.

To emphasize that the distribution of derivatives are not uncorrelated let us evaluate the fraction of points with negative potential value ($u_{0} < 0$) and positive curvature ($u_{2} > 0$)
\begin{align}\label{}
   \mathfrak{p}_{[u_{0} < 0, u_{2} > 0]} &\equiv \int_{-\infty}^{0} \!\!\! du_{0} \int_{0}^{\infty} \!\!\! du_{2}\, p(u_{0}, u_{2})
   \\ \nonumber
   &=
   \frac{1}{4} + \frac{1}{2\pi}\arctan\!\Big[ \frac{|G^{(2)}|}{\sqrt{|G^{(0)}| |G^{(4)}| - |G^{(2)}|^{2}}}\!\Big],
\end{align}
with a positive-valued argument of the $\arctan$-function. The finding $\mathfrak{p}_{[u_{0} < 0, u_{2} > 0]} > \mathfrak{p}_{[u_{0} < 0]}\mathfrak{p}_{[u_{2} > 0]}$ implies that for negatively-valued potentials, the curvature is more likely convex than concave. A similar calculation provides that the probability of finding both a positive quadratic and quartic contribution is smaller than 1/2. Curiously, this has direct implications on the properties of random Ginzburg-Landau theories: In fact, more often than not at least the sixth order contribution has to be included to guarantee the stability of the theory in the proximity of a phase transition.

The complexity of constraints can gradually be increased. For example, the \emph{perturbation-safe area fraction}, i.e., the fraction of points that are stable ($u_{2} \!>\! 0$) and robust against a perturbation of strength $E$, i.e., $E<U_{b}$, requires to choose $f(\vec{u}) = \Theta(u_{2}) \Theta[(2u_{2}^{3}/3E) - u_{3}^{2}]$, see Eq.~\eqref{eq:threshold}, with $\Theta(u)$ the Heaviside function. We have
\begin{align}\label{eq:short1}
   &\mathfrak{p}_{[u_{2} > 0,  u_{3}^{2} < 2u_{2}^{3}/3E ]} = \\
   \nonumber
   &\qquad\qquad
   \int_{-\infty}^{\infty} \!du_{2}\, du_{3}\, p(u_{2}, u_{3}) \Theta(u_{2}) \Theta[(2u_{2}^{3}/3E) \!-\! u_{3}^{2}].
\end{align}
While the quantity \eqref{eq:short1} can be evaluated for any $E$, the implicit approximation that the cubic term dominates the barrier height in Eq.~\eqref{eq:threshold} is strictly true only in the limit of small $E$, see Appendix \ref{app:truncation}.
The closed-form solution of evaluating \eqref{eq:short1} involves Euler's gamma function $\Gamma(z)$ and other special functions and is given in Appendix~\ref{app:short-to-long}. The result is a function of the dimensionless ratio $E/E_{0}$, with $E_{0} = (6 |G^{(4)}|^{3} / |G^{(6)}|^{2})^{1/2}$ defining the characteristic energy scale. For small and large values of $E/E_{0}$ the probability assumes the asymptotic forms
\begin{align}\label{eq:robust-probability}
   &\mathfrak{p}_{[u_{2} > 0,  u_{3}^{2} < 2u_{2}^{3}/3E ]} =
   \nonumber\\
   &\qquad
   =
   \left\{
   \begin{aligned}
   &\frac{1}{2}
      - \frac{\sqrt{3} \Gamma[5/6]}{2^{1/3}\pi}\Big(\frac{E}{E_{0}}\Big)^{1/3}
      &\text{for } E \ll E_{0}
      \\
      &\frac{\Gamma(3/4) \Gamma(1/12) \Gamma(5/12)}{12 \sqrt{2} \pi^{2} (E/E_{0})^{1/2}}
      &\text{for } E \gg E_{0}
   \end{aligned}
   \right.
   \\
   &\qquad\approx
   \left\{
   \begin{aligned}
   &1/2
      - 0.494 (E / E_{0})^{1/3}
      &\text{for } E \ll E_{0}
      \\
      &0.179 (E/E_{0})^{-1/2}
      &\text{for } E \gg E_{0}
   \end{aligned}
   \right.
\end{align}
In the limit $E/E_{0} \to 0$ we recover that half of the points can (upon appropriate tilt) become local minima, while the other half will be negatively curved. This area-fraction differs from the probability of finding perturbation-safe minima for \emph{a known tilt}, which is a subset of points that are minima at that tilt. For instance, for the potential at zero tilt, $u_{1} \!=\! 0$, the corresponding probability is
$\mathfrak{p}_{[u_{2} > 0, u_{3}^{2} < 2u_{2}^{3} / 3E\,|\,u_{1} = 0]}$ follows the same functional dependence as Eq.~\eqref{eq:robust-probability}, where now $|G^{(6)}|$ is replaced by $|G^{(6)}| \!-\! |G^{(4)}|^{2}/ |G^{(2)}|$. This defines a different energy scale $E_{1} = E_{0} \frac{|G^{(2)}| |G^{(6)}|}{|G^{(2)}| |G^{(6)}| - |G^{(4)}|^{2}} > E_{0}$. From this last inequality, one concludes that minima at zero tilt are more stable against perturbations than a generic minimum produced by appropriate tilting. 

\subsection{Thermal activation}
\begin{figure}[tb]
\includegraphics[width=0.4\textwidth]{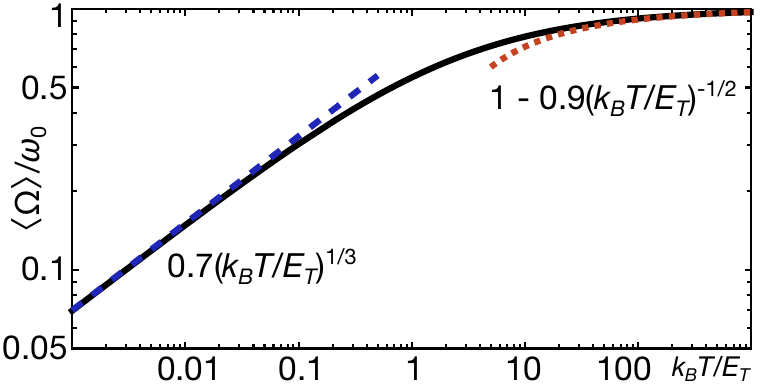}
\caption{
Temperature-dependence of the expected thermal activation frequency $\langle \Omega \rangle$ (in units of the attempt frequency $\omega_{0}$) of a particle placed in a minimum of the Gaussian random potential. Temperature is given in the characteristic unit $E_{T}/\kB$, with $E_{T} = \big(\frac{3 |G^{(2)}|^{2} |G^{(4)}|^{3}}{2 (|G^{(2)}| |G^{(6)}| - |G^{(4)}|^{2})^{2}}\big)^{1/2}$ solely given by the potential's Gaussian correlator. The activation frequency scales as $T^{1/3}$ for small temperature and approaches $\omega_{0}$ as $T^{-1/2}$ for large temperatures.}
\label{fig:T-activation}
\end{figure}

If a system is solely subjected to thermal noise the characteristic frequency for activating a particle out of a specific minimum is dictated by the barrier $U_{b}$ through Arrhenius' law
\begin{align}\label{eq:arrhenius}
   \Omega = \omega_{0} e^{- \beta U_{b}},
\end{align}
with $\beta = 1/\kB T$ and where $\omega_{0}$ is a microscopic attempt frequency. For a particle, randomly placed in a potential minimum (at zero tilt), the characteristic frequency for escape becomes
\begin{align}
   \langle \Omega \rangle &= \omega_{0} \sqrt{8\pi |G^{(2)}|}\; \times \nonumber\\
   \label{eq:short2}
   &\quad\int_{0}^{\infty} \!\!\!du_{2} \int_{-\infty}^{\infty} \!\!\! du_{3} p(u_{1} \!=\! 0,u_{2},u_{3}) e^{- \beta 2u_{2}^{3} / 3u_{3}^{2}}.
\end{align}
Here we have assumed again, that the barrier is appropriately described by Eq.~\eqref{eq:threshold}, see discussion in Appendix~\ref{app:truncation}. The factor $\mathfrak{p}_{[u_{1} = 0, u_{2} > 0]} = (8\pi |G^{(2)}|)^{-1/2}$ guarantees the appropriate normalization by imposing that the particle initially occupies a minimum at zero tilt. While the full form of the result is again given in Appendix~\ref{app:short-to-long}, its asymptotic behavior is
\begin{align}
   \!\!
   \langle \Omega \rangle(\beta) &=
   \!\left\{\!
   \begin{aligned}
   &\omega_{0} \frac{2^{2/3} \Gamma(\frac{2}{3})}{(3\pi)^{1/2}} (\beta E_{T})^{-1/3} &\!\!\!\!\text{for } \beta E_{T} \!\gg\! 1
   \\
   &\omega_{0} \Big[1 \!-\! \frac{\Gamma(\frac{1}{12}) \Gamma(\frac{5}{12}) \Gamma(\frac{3}{4})}{6 \pi^{3/2}} \sqrt{\beta E_{T}} \Big] &\!\!\!\!\text{for } \beta E_{T} \!\ll\! 1
   \end{aligned}
   \right.\!\!
\end{align}
Here $E_{T} = \big(\frac{3 |G^{(2)}|^{2} |G^{(4)}|^{3}}{2 (|G^{(2)}| |G^{(6)}| - |G^{(4)}|^{2})^{2}}\big)^{1/2}$ is the characteristic thermal energy above which the system is subject to full thermal unrest. As might be expected, it is only numerically different from $E_{0}$ defined above. The scaling of the thermal activation frequency with temperature---$T^{1/3}$ ($\sim1 -T^{-1/2}$) at low (high) temperature---provides direct access to the potential characteristics defining $E_{T}$. While the expansion for large temperatures, $\beta E_{T} \ll 1$, is formally correct, the earlier approximation that the potential expansion can be truncated after the third order is violated. Similar to the previous section \ref{sec:area-fractions}, this calculation can be repeated for different tilts $u_{1}$ which allows to extract all three parameters $G^{(2)}$, $G^{(4)}$, and $G^{(6)}$ independently, at least in principle.

To observe the predicted $T^{1/3}$ low-temperature scaling of $\langle \Omega \rangle$ experimentally, we recall the physics of individual superconducting vortices trapped by material defects. In scanning probe setups such as SQUID-on-Tip or STM creep spectroscopy, a single vortex is held in a metastable pinscape where the attempt frequency $\omega_0$ is governed by the vortex core's depinning resonance, typically in the gigahertz range. By tracking the temperature dependence of the mean escape rate $\langle \Omega \rangle(T)$ under vanishing applied tilt ($u_1 = 0$), the low-temperature slope directly reveals the characteristic thermal crossover energy $E_T$. Because $E_T$ only depends on the even derivatives $G^{(2)}$, $G^{(4)}$, and $G^{(6)}$, combining this escape-rate spectroscopy with measurements of the stable area fraction ($p_{u_2>0}$) allows one to infer the variance of the pinning force of the underlying disorder landscape.

\subsection{Quantum tunneling}

\begin{figure}[tb]
\includegraphics[width=0.4\textwidth]{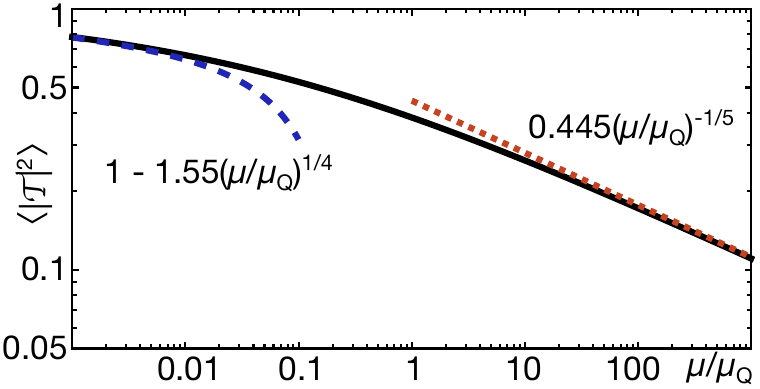}
\caption{
Mass-dependent probability $\langle |\mathcal{T}|^{2}\rangle$ for a quantum particle to tunnel away from a metastable minimum of a Gaussian random potential. The particle's mass $m$ is given in units of $\hbar^{2} \mu_{Q}/2$, where $\mu_{Q} \equiv \big(\frac{512 (|G^{(2)}| |G^{(6)}| - |G^{(4)}|^{2})^{4}}{405 |G^{(2)}|^{4} |G^{(4)}|^{5}}\big)^{1/2}$ depends on the two-point correlator of the potential. The tunneling probability decreases as $m^{-1/5}$ for large masses and approaches unity as $m^{1/4}$ in the limit $m \to 0$.}
\label{fig:Q-activation}
\end{figure}
We conclude the section with evaluating the quantum tunneling rate for a particle found at rest in a local minimum of the random potential. Within the WKB approximation, the characteristic probability for tunneling into a neighboring minimum is given by
\begin{align}\label{eq:WKB}
   |\mathcal{T}|^{2} \sim e^{ - 2 \Lambda_{b}}
\end{align}
where $\Lambda_{b} \equiv \int_{b}\! dx \sqrt{(2m/\hbar^{2})[U(x) - U(0)]}$ the integral over the potential barrier and $m$ the particle mass. In our case, we find $\Lambda_{b}  = \sqrt{2 (2m/\hbar^{2})}(6 u_{2}^{5/2} / 5u_{3}^{2})$. Averaging over all possible realizations, one finds the characteristic tunneling rate is
\begin{align}\label{eq:short3}
   \langle |\mathcal{T}|^{2} \rangle
   &\!\sim\!\! \int_{0}^{\infty} \!\!\!\!du_{2} \int_{-\infty}^{\infty} \!\!\!\! du_{3} \frac{p(u_{1} \!=\! 0,u_{2},u_{3})}{\mathfrak{p}_{[u_{1} = 0, u_{2} > 0]}} e^{- 2 \sqrt{8 \mu u_{2}^{5}} / 15u_{3}^2},
\end{align}
with $\mu \equiv 2m / \hbar^{2}$ a measure of the effective mass. As before, this expression can be cast in the form of generalized hypergeometric functions, as shown in Appendix~\ref{app:short-to-long}, and is a dimensionless function of $\mu/\mu_{Q}$, with $\mu_{Q} \equiv \big(\frac{512 (|G^{(2)}| |G^{(6)}| - |G^{(4)}|^{2})^{4}}{405 |G^{(2)}|^{4} |G^{(4)}|^{5}}\big)^{1/2}$. For large masses, $\mu \gg \mu_{Q}$ the tunneling probability decays as a power law
\begin{align}
   \langle |\mathcal{T}|^{2} \rangle(\mu) \sim \nu_{>} \, (\mu / \mu_{Q})^{-1/5}
\end{align}
with $\nu_{>} \approx0.445$.
For small masses $\mu \ll \mu_{Q}$ the tunneling probability instead saturates as
\begin{align}
   \langle |\mathcal{T}|^{2} \rangle(\mu) \sim 1 - \nu_{<}\, (\mu / \mu_{Q})^{1/4}
\end{align}
with $\nu_{<} \approx 1.55$. Although closed expression for $\nu_{\lessgtr}$ can be derived, they do not simplify to a reasonable form and are not given here. Similar to the thermal case, the limiting case $\mu \ll \mu_{Q}$ will be modified by higher order contributions in the potential expansion.

\section{Numerics}\label{sec:numerics}
\begin{figure}
\includegraphics[width=0.65\textwidth]{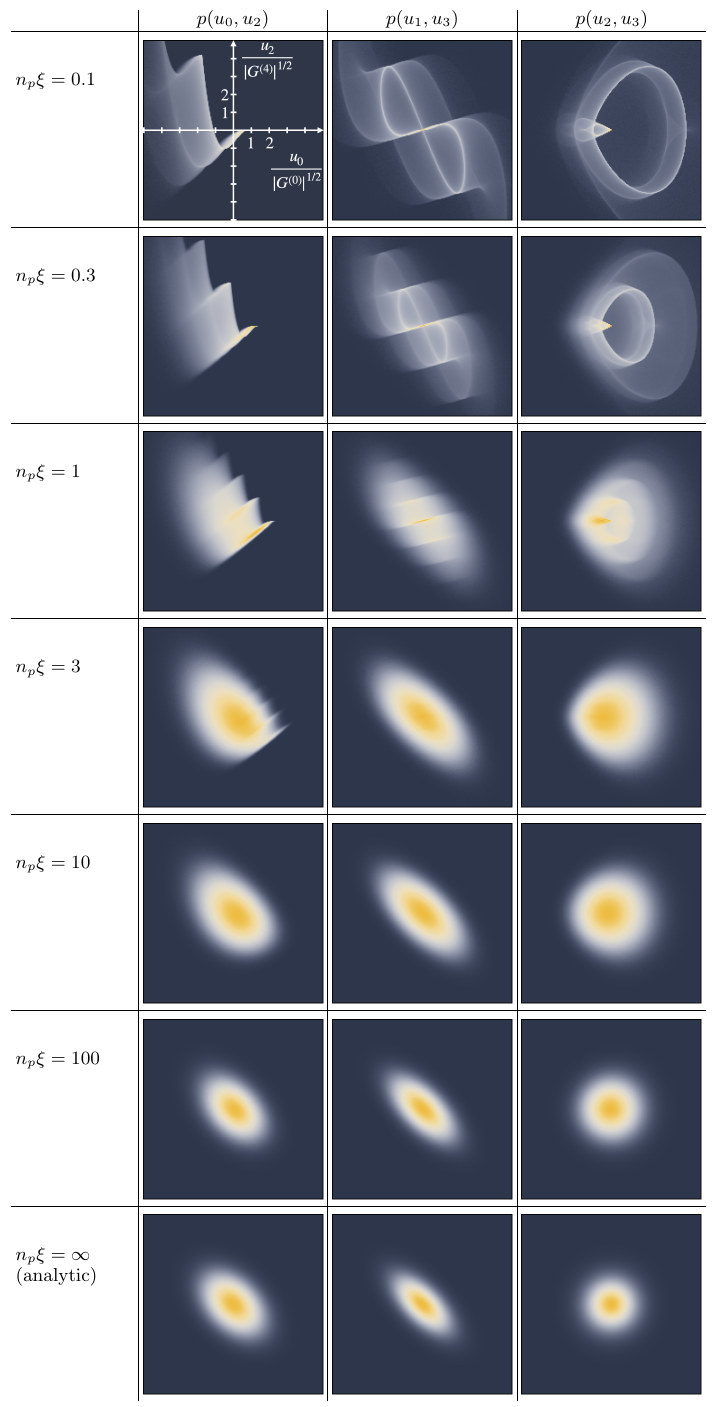}
\caption{
Probability distributions $p(u_{0}, u_{2})$, $p(u_{1}, u_{3})$, $p(u_{2}, u_{3})$ for different densities $n_{p} = N_{p} / L$ of Lorentzian defects, in a system of length $L = 100\xi$, see Eq.~\eqref{eq:Lorenzian}.
The plot range $\pm 5\sqrt{|G^{(2n)}|}$ scales with the defect density as $n_{p}^{1/2}$. For $n_{p}\xi > 1$, the paired probability distributions rapidly converge towards the analytic result valid for $n_{p}\xi \to \infty$ (central/Gaussian limit).
While $p(u_{0}, u_{2})$ and $p(u_{1}, u_{3})$ are correlated (tilted ellipses), the second and third derivatives are decoupled $p(u_{2}, u_{3}) = p(u_{2}) p(u_{3})$ in the Gaussian limit.
On a technical note: We compute $N_{r}$ potential realizations and deduce from it a histogram $h(u_{i}, u_{j})$ related to the probability distribution $p(u_{i}, u_{j})$ via $p(u_{i}, u_{j}) = \lim_{N_{r}\to \infty}[h(u_{i}, u_{j})/N_{r}]$. The color-scaling function $\ln[h(u_{i}, u_{j}) + N_{p}]$ has turned out to provide a good visual contrast in the plots when varying the defect density over several orders of magnitude.
}
\label{fig:densityplot}
\end{figure}

In the analytic derivations above we have computed the probability distribution $p(\vec{u})$ for a Gaussian random potential. It has also been shown in Ref.~\cite{Willa2022} that this limit is reached for a random potential produced by a large density $n_{p}$ of (smooth) defects, i.e. in the limit $n_{p} \xi \gg 1$ with $\xi$ the characteristic size of the defect. For practical purposes, it is interesting to study how this limit is reached and---in view of the above considerations of computing $\langle \Omega \rangle$ and $\langle |\mathcal{T}|^{2} \rangle$---how the probability distributions $p(u_{0}, u_{2})$, $p(u_{1}, u_{3})$, and $p(u_{2}, u_{3})$ converge to the Gaussian limit.
To this end we compute the characteristic properties of a random potential landscape numerically. Practically, a system is populated with $N_{p}$ identical Lorentzian potential wells
\begin{align}\label{eq:Lorenzian}
   V(x - x_{\ell}) \equiv \frac{-V_{0}}{1 + (x - x_{\ell})^2/\xi^{2}} + \bar{V}
\end{align}
at random positions $-L/2 < x_{\ell} < L/2$. Here, $L = 100\xi$ is the system size while $\xi$ and $V_{0}$ are the natural length and energy scales of the defect. The constant $\bar{V} = 2 (V_{0}\xi/L)  \arctan(L/2\xi)\approx \pi V_{0}\xi/L$ guarantees that the mean potential value vanishes. For one such realization, the potential derivatives $U^{(n)}(0)$ [$n \in \{0,1,2,3\}$] are evaluated. The system's center suffers least from boundary effects. Upon repeating this process for up to $10^{8}$ potential realizations, we evaluate the probability distribution for these quantities, as shown in Figure~\ref{fig:densityplot}.

For this potential the two-point correlator can be calculated explicitly, yielding
\begin{align}\label{}
   G_{L}(x) = n_{p}\xi \frac{2 \pi V_{0}^{2}}{4 + x^{2}/\xi^{2}}
\end{align}
with $n_{p} \equiv N_{p}/L$ the defect density. We have $G_{L}(0) = (\pi n_{p} \xi V_{0}^{2})/2$, even (non-zero) derivatives take the universal form
\begin{align}\label{eq:G2m}
   G_{L}^{(2m)}(0) = \frac{(-1)^{m} 2m\, \Gamma(2m)}{4^{m} \xi^{2m}} G_{L}(0),
\end{align}
and all odd derivatives vanish at the origin. 
Figure~\ref{fig:densityplot} displays the probability density of pairs of potential derivatives, namely $p(u_{0}, u_{2})$, $p(u_{1}, u_{3})$, as well as $p(u_{2}, u_{3})$ [needed for computing the above observables]. These probability densities are shown for finite densities $n_{p}\xi = N_{p} \xi / L$ in the range 0.1-100 and the analytic dependence is given in the last row. An animation of the evolution for intermediate densities is provided as a supplementary material \cite{supp-mat-movies}.

\begin{acknowledgments}
We cordially thank G.\ Blatter, Y.\ Fyodorov, V.B.\ Geshkenbein, A.\ Mirlin, C.\ Sp\r{a}nsl\"{a}tt for enlightening discussions.
R.W. acknowledges the support from the Heidelberger Akademie der Wissenschaften (WIN, 8. Teilprogramm).
\end{acknowledgments}

\twocolumngrid
\bibliographystyle{apsrev4-1-titles}
\bibliography{StabilityInRandomPotentials}

\vfill
\onecolumngrid

\vfill
\pagebreak
\quad

\appendix
\onecolumngrid
\section{Full analytic solutions}
\label{app:short-to-long}

The full expressions of the computed results \eqref{eq:short1}, \eqref{eq:short2}, \eqref{eq:short3} require Euler's gamma function $\Gamma(s) \equiv \int_{0}^{\infty}d\zeta\, \zeta^{s-1}e^{-\zeta}$ and the generalized hypergeometric function ${}_{p}F_{q}\big[\mNoBracket{a_{1},\\ b_{1},}\ \mNoBracket{..., \\ ...,}\ \mNoBracket{a_{p}  \\ b_{q} }; z \big] \equiv \sum_{n=0}^{\infty}\frac{(a_{1})_{n} \cdots (a_{p})_{n}}{(b_{1})_{n} \cdots (b_{q})_{n}} \frac{z^{n}}{n!}$ [with $(c)_{n} \equiv c(c \!+\! 1)\dots(c \!+\! n \!-\! 1)$ [$(c)_{0} \equiv 1$] the Pochhammer symbol] and are given here
\begin{align}\label{eq:long1}
   &\mathfrak{p}_{[u_{2} > 0,  u_{3}^{2} < 2u_{2}^{3}/3E ]}
      = \int_{-\infty}^{\infty} du_{2}\, du_{3}\, p(u_{2}, u_{3}) \Theta(u_{2}) \Theta[(2u_{2}^{3}/3E) - u_{3}^{2}]
      = \int_{0}^{\infty} du_{2}\, p(u_{2}) \Erf\Big[\sqrt{\frac{u_{2}^{3} }{3 E |G^{(6)}|}}\Big]
      \\
      &\qquad
      = \frac{1}{2}
      - \frac{\sqrt{3} (E/E_{0})^{1/3} \Gamma[5/6]}{2^{1/3}\pi}\;
        {}_{3}F_{3}\Big[\mNoBracket{1/6,  \\ 1/3,}\ 
                        \mNoBracket{5/12, \\ 2/3,}\ 
                        \mNoBracket{11/12 \\ 7/6 };
                        -(E/E_{0})^{2}\Big]
      \\\nonumber
      &\qquad\qquad
      + \frac{\sqrt{3} (E/E_{0})}{4 \sqrt{\pi}}\;
        {}_{3}F_{3}\Big[\mNoBracket{1/2, \\ 2/3,}\ 
                        \mNoBracket{3/4, \\ 4/3,}\ 
                        \mNoBracket{5/4  \\ 3/2 };
                        -(E/E_{0})^{2}\Big]
      - \frac{9 \sqrt{3} (E/E_{0})^{5/3} \Gamma[13/6]}{20\; (2^{2/3}) \pi}\;
        {}_{3}F_{3}\Big[\mNoBracket{5/6,   \\ 4/3,}\ 
                        \mNoBracket{13/12, \\ 5/3,}\ 
                        \mNoBracket{19/12  \\ 11/6 };
                        -(E/E_{0})^{2}\Big],
\end{align}
\begin{align}\label{}\label{eq:long2}
   \langle \Omega \rangle &= \int_{0}^{\infty} \!\!\!du_{2} \int_{-\infty}^{\infty} \!\!\! du_{3} \frac{p(u_{1} \!=\! 0,u_{2},u_{3})}{\mathfrak{p}_{[u_{1} = 0, u_{2} > 0]}} \omega_{0} e^{- \beta 2u_{2}^{3} / 3u_{3}^{2}}
      \\
      &=
      \omega_{0}\Big\{
      {}_{2}F_{2}\Big[\mNoBracket{1/6, \\ 1/4,}\ 
                      \mNoBracket{5/6 \\ 3/4};
                      (\beta E_{T})^{2}\Big]
      -\frac{2 \Gamma(5/12) \Gamma(3/4) \Gamma(13/12) (\beta E_{T})^{1/2}}{\pi^{3/2}}
      {}_{2}F_{2}\Big[\mNoBracket{ 5/12, \\ 1/2,}\ 
                      \mNoBracket{13/12 \\ 5/4};
                      (\beta E_{T})^{2}\Big]
      \\\nonumber
      &\qquad
      +\frac{8 (\beta E_{T})}{3 (3\pi)^{1/2}}
      {}_{3}F_{3}\Big[\mNoBracket{2/3, \\ 3/4,}\ 
                      \mNoBracket{  1, \\ 5/4,}\ 
                      \mNoBracket{4/3 \\ 3/2};
                      (\beta E_{T})^{2}\Big]
      -\frac{16 \Gamma(11/12) \Gamma(5/4) \Gamma(19/12) (\beta E_{T})^{3/2}}{3 \pi^{3/2}}
      {}_{2}F_{2}\Big[\mNoBracket{11/12, \\ 3/2,}\ 
                      \mNoBracket{19/12 \\ 7/4};
                      (\beta E_{T})^{2}\Big]
      \Big\},
\end{align}
\begin{align}\label{eq:long3}
   &\langle |\mathcal{T}|^{2} \rangle \sim \int_{0}^{\infty} \!\!\!du_{2} \int_{-\infty}^{\infty} \!\!\! du_{3} \frac{p(u_{1} \!=\! 0,u_{2},u_{3})}{\mathfrak{p}_{[u_{1} = 0, u_{2} > 0]}} e^{- (12/5) \sqrt{2 \mu  u_2^{5}} / u_{3}^2} = 
      \\
      &\quad
      {}_{4}F_{6}\Big[\mNoBracket{1/10, \\ 1/8,}\
                      \mNoBracket{3/10, \\ 2/8,}\
                      \mNoBracket{7/10, \\ 3/8,}\
                      \mNoBracket{9/10  \\ 5/8,}\
                      \mNoBracket{      \\ 6/8,}\
                      \mNoBracket{      \\ 7/8};
                      \Big(\frac{\mu}{\mu_{Q}}\Big)^{\!2}\Big]
      \\ \nonumber
      &\quad
      -\frac{2 \Gamma(9/40) \Gamma(17/40) \Gamma(25/40) \Gamma(33/40) \Gamma(41/40)} {\pi^{5/2}}
      \Big(\frac{\mu}{\mu_{Q}}\Big)^{\!1/4}
      {}_{4}F_{6}\Big[\mNoBracket{ 9/40, \\ 2/8,}\
                      \mNoBracket{17/40, \\ 3/8,}\
                      \mNoBracket{33/40, \\ 4/8,}\
                      \mNoBracket{41/40  \\ 6/8,}\
                      \mNoBracket{       \\ 7/8,}\
                      \mNoBracket{       \\ 9/8};
                      \Big(\frac{\mu}{\mu_{Q}}\Big)^{\!2}\Big]
      \\ \nonumber
      &\quad
      +\frac{8 \Gamma(7/20) \Gamma(11/20) \Gamma(15/20) \Gamma(19/20) \Gamma(23/20)} {\pi^{5/2}}
      \Big(\frac{\mu}{\mu_{Q}}\Big)^{\!1/2}
      {}_{4}F_{6}\Big[\mNoBracket{ 7/20, \\ 3/8,}\
                      \mNoBracket{11/20, \\ 4/8,}\
                      \mNoBracket{19/20, \\ 5/8,}\
                      \mNoBracket{23/20  \\ 7/8,}\
                      \mNoBracket{       \\ 8/8,}\
                      \mNoBracket{       \\ 10/8};
                      \Big(\frac{\mu}{\mu_{Q}}\Big)^{\!2}\Big]
      \\ \nonumber
      &\quad
      -\frac{64 \Gamma(19/40) \Gamma(27/40) \Gamma(35/40) \Gamma(43/40) \Gamma(51/40)} {3\pi^{5/2}}
      \Big(\frac{\mu}{\mu_{Q}}\Big)^{\!3/4}
      {}_{4}F_{6}\Big[\mNoBracket{19/40, \\ 4/8,}\
                      \mNoBracket{27/40, \\ 5/8,}\
                      \mNoBracket{43/40, \\ 6/8,}\
                      \mNoBracket{51/40  \\ 9/8,}\
                      \mNoBracket{       \\ 10/8,}\
                      \mNoBracket{       \\ 11/8};
                      \Big(\frac{\mu}{\mu_{Q}}\Big)^{\!2}\Big]
      \\ \nonumber
      &\quad
      +\frac{1024}{75 (5\pi)^{1/2}}
      \Big(\frac{\mu}{\mu_{Q}}\Big)
      {}_{5}F_{7}\Big[\mNoBracket{3/5, \\ 5/8,}\
                      \mNoBracket{4/5, \\ 6/8,}\
                      \mNoBracket{1,   \\ 7/8,}\
                      \mNoBracket{6/5, \\ 9/8,}\
                      \mNoBracket{7/5  \\ 10/8,}\
                      \mNoBracket{     \\ 11/8,}\
                      \mNoBracket{     \\ 12/8};
                      \Big(\frac{\mu}{\mu_{Q}}\Big)^{\!2}\Big]
      \\ \nonumber
      &\quad
      -\frac{1024 \Gamma(29/40) \Gamma(37/40) \Gamma(45/40) \Gamma(53/40) \Gamma(61/40)} {15\pi^{5/2}}
      \Big(\frac{\mu}{\mu_{Q}}\Big)^{\!5/4}
      {}_{4}F_{6}\Big[\mNoBracket{29/40, \\ 6/8,}\
                      \mNoBracket{37/40, \\ 7/8,}\
                      \mNoBracket{53/40, \\ 10/8,}\
                      \mNoBracket{61/40  \\ 11/8,}\
                      \mNoBracket{       \\ 12/8,}\
                      \mNoBracket{       \\ 13/8};
                      \Big(\frac{\mu}{\mu_{Q}}\Big)^{\!2}\Big]
      \\ \nonumber
      &\quad
      +\frac{4096 \Gamma(17/20) \Gamma(21/20) \Gamma(25/20) \Gamma(29/20) \Gamma(33/20)} {45 \pi^{5/2}}
      \Big(\frac{\mu}{\mu_{Q}}\Big)^{\!3/2}
      {}_{4}F_{6}\Big[\mNoBracket{17/20, \\ 7/8,}\
                      \mNoBracket{21/20, \\ 9/8,}\
                      \mNoBracket{29/20, \\ 11/8,}\
                      \mNoBracket{33/20  \\ 12/8,}\
                      \mNoBracket{       \\ 13/8,}\
                      \mNoBracket{       \\ 14/8};
                      \Big(\frac{\mu}{\mu_{Q}}\Big)^{\!2}\Big]
      \\ \nonumber
      &\quad
      -\frac{32768 \Gamma(39/40) \Gamma(47/40) \Gamma(55/40) \Gamma(63/40) \Gamma(71/40)} {315\pi^{5/2}}
      \Big(\frac{\mu}{\mu_{Q}}\Big)^{\!7/4}
      {}_{4}F_{6}\Big[\mNoBracket{39/40, \\ 9/8,}\
                      \mNoBracket{47/40, \\ 10/8,}\
                      \mNoBracket{63/40, \\ 12/8,}\
                      \mNoBracket{71/40  \\ 13/8,}\
                      \mNoBracket{       \\ 14/8,}\
                      \mNoBracket{       \\ 15/8};
                      \Big(\frac{\mu}{\mu_{Q}}\Big)^{\!2}\Big].
\end{align}

\section{Truncation after third order}\label{app:truncation}
The result in Eq.~\eqref{eq:G2m} allows one to evaluate the relevance of higher order derivatives in characterizing the instability point. The barrier produced by the $n$-th potential term is obtained from evaluating the extrema of
\begin{align}\label{}
   U_{n}(x) = u_{2} x^{2} / 2 + u_{n} x^{n} / n!
\end{align}
and scales as
\begin{align}\label{eq:barrier-height}
   U_{b,n} = \frac{n - 2}{2n} \Big[\frac{ u_{2} (n \!-\! 1)!}{ u_{n}}\Big]^{\frac{2}{n-2}} u_{2}.
\end{align}
Inserting the characteristic values $u_{n} \sim \sqrt{|G^{(2n)}|}$ into the above expression, we find that the square bracket is of order unity [with a logarithmic dependence $\ln(n)/n$ for large $n$]. Both the barrier height \eqref{eq:barrier-height} and its position
\begin{align}\label{}
   x_{b,n} = \Big[\frac{- u_{2} (n \!-\! 1)!}{ u_{n}}\Big]^{\frac{1}{n-2}}
\end{align}
away from the minimum are growing with increasing $n$. Despite the absence of a parametric difference, this trend justifies the truncation of the expansion in determining the barrier \eqref{eq:threshold} for \emph{small} perturbations around the equilibrium position. In fact, mostly small barriers contribute to the escape rates, $\langle\Omega\rangle$ \eqref{eq:arrhenius} and $\langle|\mathbb{\mathcal{T}}|^{2}\rangle$ \eqref{eq:WKB} and justifies the results obtained above

For the thermally activated escape it is possible to define a characteristic energy $\bar{E}$ according to Arrhenius' law
\begin{align}
\langle\Omega\rangle=\omega_{0}e^{-\beta\bar{E}}\Leftrightarrow\bar{E}=k_{B}T\ln(\omega_{0}/\langle\Omega\rangle)
\end{align}
showing mainly barriers with $U_{b}<\bar{E}\approx k_{B}T$ contribute. At temperatures $k_{B}T<U_{b,4}$ a truncation of the potential expansion after the third order is justified. At higher temperatures larger barriers also have to be included and the calculated rate $\langle\Omega\rangle$ becomes an upper limit for an exact result.

Similarly, for the escape by quantum tunneling we define a barrier $\bar{U}(r)$ with $\bar{\Lambda}_{b}$ like in Eq. \eqref{eq:WKB} satisfying
\begin{align}
\langle|\mathcal{T}|^{2}\rangle\sim e^{-2\bar{\Lambda}_{b}}\Leftrightarrow\bar{\Lambda}_{b}\sim-\ln(\langle|\mathcal{T}|^{2}\rangle)
\end{align}
to show the barriers with $\Lambda_{b}<\bar{\Lambda}_{b}$ contribute the most where $\bar{\Lambda}_{b}$ is of order unity. Therefore masses $m$ where $m>\hbar^{2}[\int_{b}dx\sqrt{U_{4}(x)-U(0)}]^{-2}$ justify truncating the expansion after the third order. For smaller masses the calculated tunneling rate $\langle|\mathcal{T}|^{2}\rangle$ provides an upper boundary to the exact result.
\end{document}